\documentclass[preprint,floatfix] {revtex4} 
 
\usepackage{graphicx}
\usepackage{subfigure}
\usepackage{amsmath}
\begin{document}

%234567890123456789012345678901234567890123456789012345678901234567890123456789012345678901234567890
\title{Spherical confinement of Coulombic systems inside an impenetrable box: H atom and the Hulth\'en potential}
\author{Amlan K. Roy}
\altaffiliation{Email: akroy@iiserkol.ac.in, akroy6k@gmail.com, Ph: +91-3473-279137, Fax: +91-33-25873020.}
\affiliation{Division of Chemical Sciences, \\   
Indian Institute of Science Education and Research Kolkata, 
Mohanpur Campus, Nadia, 741252, India}

\begin{abstract}
The generalized pseudospectral method is employed to study spherical confinement in two simple Coulombic systems: (i) well celebrated and 
heavily studied H atom (ii) relatively less explored Hulth\'en potential. In both instances, \emph{arbitrary} cavity size, as well as \emph{low and 
higher states} are considered. Apart from bound state eigenvalues, eigenfunctions, expectation values, quite accurate estimates of the critical
cage radius for H atom for all the 55 states corresponding to $n \leq 10$, are also examined. Some of the latter are better than previously 
reported values. Degeneracy and energy ordering under the isotropic confinement situation are discussed as well. The method
produces consistently high-quality results for both potentials for \emph{small as well as large} cavity size. For the H atom, present results are 
comparable to best theoretical values, while for the latter, this work gives considerably better estimates than all existing work
so far. 

\vspace{0.5in}

\end{abstract}
\maketitle

\section{Introduction}
Enclosure of an atom/molecule in a spherically impenetrable box was first conceived as early as in 1937 \cite{michels37}, where  
the effects of high pressure on energy levels, polarizability and ionization potential were studied. A many-electron system trapped 
inside an inert cavity with such a boundary experiences spatial confinement that affects its physical and chemical properties. 
This has, therefore, been employed in a variety of situations, e.g., the cell-model of liquid state, high-pressure physics, 
study of impurities in semiconductor materials, matrix isolated molecules, endohedral complexes of fullerenes, zeolites cages, 
helium droplets, nano-bubbles, etc.~\cite{fernandez82a,froman87,jaskolski96,connerade00,buchachenko01,gravesen05,heiss05}. This has also found 
astrophysical applications, such as hydrogen atom spectra \cite{sommerfeld38}, mass-radius relation in theory of white dwarfs, 
determination of rate of escape of stars from galactic and globular clusters, simulation of the interiors of giant planets Jupiter 
and Saturn \cite{guillot99}, etc. The recent upsurge of interest in nanotechnology has also inspired extensive research activity to 
simulate spatially confined quantum systems (on a scale comparable to their de Broglie wave length). Importance of such 
\emph{artificial} atoms has been realized in quantum wells, quantum wires, quantum dots as well as nano-sized circuits such 
as quantum computer, etc., by employing a wide variety of confining potentials. 

Ever since the pioneering model of \cite{michels37} on compressed quantum systems, an enormous amount of work has been reported on 
confined hydrogen atom (CHA) problem, in particular. Effect of isotropic compression on $1s$, $2s$ and $2p$ levels of H atom were 
provided through a semi-quantitative calculation \cite{degroot46}. These \cite{michels37, degroot46} and other following work
\cite{suryanarayana76} invoked a direct solution of relevant Schr\"odinger equation imposing the boundary condition that wave 
function vanishes at the surface of enclosing sphere. A Hartree-Fock self-consistent field solution \cite{ludena77} with Slater-type 
orbitals and cut-off functions has been proposed. Approximate analytical formulas for CHA eigenvalues were derived using Vawter's $\coth z$ method 
\cite{vawter73}, joint perturbation method and Pad\'e approximation \cite{navarro80, arteca83}, WKB method \cite{sinha03}. 
Some other attempts are: a combined hyper-virial theorem and perturbation theory \cite{fernandez82}, a variational 
boundary perturbation method with appropriate cut-off function \cite{marin91, marin91a} along with its variants \cite{varshni97,varshni99}, 
by extending a power-series solution \cite{palma87}, originally proposed for free quantum systems, to confined case \cite{aquino95}, 
self-consistent solution \cite{garza98} of relevant Kohn-Sham equation within the 
broad domain of density functional theory, variational perturbation theory \cite{montgomery01}, variational method in 
conjunction with super-symmetric quantum mechanics \cite{filho02, filho03}, Rayleigh-Schr\"odinger perturbation theory \cite{laughlin02}, 
Lie algebraic treatment \cite{laughlin02}, Lagrange-mesh method \cite{baye08}, searching the zeros of hyper-geometric function 
\cite{shaqqor09}, asymptotic iteration method \cite{ciftci09}, etc. So far the most accurate calculations are those based on formal 
solution of confluent hyper-geometric function and series method \cite{aquino07}. Exact solutions for this system are expressed directly in terms of 
Kummer $M-$function (confluent hyper-geometric) \cite{burrows06}. Some numerical schemes have also been proposed, e.g., \cite{goldman92}. While the 
dependence of ground- and excited-state energies on cage radius remained in the center of investigation in all these 
mentioned works, a host of other properties have all also found attention. A few notable ones are: hyperfine splitting constant \cite{ludena77,
leykoo79,arteca83,aquino95,aquino07}, dipole shielding factor \cite{fowler84}, nuclear magnetic screening constant \cite{leykoo79,aquino95,
aquino07}, pressure \cite{leykoo79,arteca83,aquino95,aquino07}, excited-state life time \cite{goldman92}, nuclear volume isotope effect 
\cite{goldman92}, density derivatives at the nucleus \cite{montgomery09} etc. A significant amount of work exists on static and
dynamic polarizability \cite{leykoo79, fowler84,aquino95, dutt01, patil02, sen02, montgomery02, laughlin04, burrows05,aquino07, cohen08, cakir13} 
within the Kirkwood, Buckingham, Unsold approximations and many other methods as well. It offers some interesting properties  
in higher dimensions \cite{shaqqor09} as well. Confinement within \emph{penetrable} boundaries \cite{leykoo79,gorecki87,marin92}
are studied. Shannon and Fisher information entropies in position and momentum space of CHA in soft as well as hard spherical cavities have 
been investigated \cite{sen05, aquino13}. Various scaling relations are proposed \cite{varshni02}. Another interesting 
aspect of this problem is that as the confining radius decreases, binding energy decreases and there exists a \emph{critical} value 
of this radius ($r_c$), at which latter becomes zero. Many attempts have been made to estimate this \cite{varshni98}. Numerous other features 
of CHA as well as the methods employed could be found in the elegant reviews \cite{sen09,aquino09,aquino14} and references therein.

This work is concerned with the spherical confinement inside an impenetrable cavity of two widely used Coulombic systems, \emph{viz.,} 
H atom and Hulth\'en potential. For this, the generalized pseudospectral (GPS) method is invoked, which has been found to be quite successful for 
a number of problems \cite{roy04,roy04a,roy04b,roy05,roy05a,sen06,roy08,roy14}. However, its application to confinement situations has rather been 
limited: H atom and Davidson oscillator \cite{sen06}, 3D polynomial oscillator including the harmonic oscillator \cite{roy14}. 
Although a detailed study was made in the latter case, for H atom, only a few $s$ and $p$ states (a total of 9) were reported. Given the success
of this approach for bound states of a variety of problems (as given in the references and therein), it is desirable to assess and validate
its performance for other relevant confinement studies. With this in mind, here we thus present its extension in the context of confined H atom 
case in terms of eigenvalues, eigenfunctions, radial densities and various expectation values. Besides, the critical box radius, $r_c^c$, 
for all the 55 states in H atom, are given. Small, medium and large box sizes have been used. Energy variations with respect to the cage radius, 
$r_c$, are followed for \emph{low and high} excited states. Moreover, while a vast amount of work is published for confinement in H atom, much 
lesser attempts are known to understand its effects on other Coulombic systems. To follow this, the case of Hulth\'en potential is considered, which 
represents an important short-range potential. Applications are found in particle physics, atomic physics, solid-state physics and chemical physics
(see, for example, the references \cite{olson78,durand81,lindhard86,bechler88,bitensky97} and therein). 
Over the years, numerous methods have been proposed for accurate estimation of its bound states
in the \emph{free} system. Some scattered works have been published for this potential under spherical confinement as well \cite{sinha03,filho03}. 
We offer accurate bound-state energies of confined Hulth\'{e}n potential for ground and some low-lying states for varying range of screening parameter. 
Small as well as large $r_c$ has been considered in a systematic manner. The article is organized as follows: Section II gives a brief outline of our 
method, a discussion of the results are presented in Section III, while some concluding remarks are offered in Section IV.  

\section{The GPS method for Confinement}
Various features of the methodology were discussed detail in previous references \cite{roy04,roy04a,roy04b,roy05,roy05a,sen06,roy08,roy14}. 
Here, only the essential details, necessary for solution of relevant single-particle Schr\"odinger equation for a central potential under the 
influence of a spherical confinement, are presented. One seeks the solution of following time-independent non-relativistic eigenvalue equation:
\begin{equation}
\left[-\frac{1}{2} \ \frac{d^2}{dr^2} + \frac{\ell (\ell+1)} {2r^2} + v(r) \right] \psi_{n,\ell}(r)=E_{n,\ell}\ \psi_{n,\ell}(r),
\end{equation}
where $v(r)$ characterizes the specific potential under investigation, whereas $n, \ell$ refer to the usual radial and angular quantum 
numbers. Our interest lies in the following two cases, 
\begin{equation} v(r)= \begin{cases} 
-\frac{1}{r}, \ \ \ \ \ \ \ \ \ \ \mathrm{for} \  \mathrm{H} \ \mathrm{atom} \\
-\frac{\delta e^{-\delta r}}{1-e^{-\delta r}}, \ \ \ \  \mathrm{for} \ \mathrm{Hulth\acute{e}n} \ \mathrm{potential}, \\
\end{cases}
\end{equation} 
where $\delta$ is a screening parameter. 
The spherical confinement is achieved by introducing the following potential as $v_c(r) = +\infty$ for $r > r_c$ and 0 for $r \leq r_c$, 
where $r_c$ signifies the radius of spherical enclosure. One needs to solve this equation satisfying the Dirichlet boundary condition, 
$\psi_{n, \ell}\ (0)=\psi_{n,\ell} \ (r_c)=0$.

The crucial step is to approximate a function $f(x)$ defined in the interval $x \in [-1,1]$ by an N-th order polynomial $f_N(x)$,
\begin{equation}
f(x) \cong f_N(x) = \sum_{j=0}^{N} f(x_j)\ g_j(x),
\end{equation}
so that the approximation is \emph {exact} at \emph {collocation points} $x_j$, i.e., $ f_N(x_j) = f(x_j).$
Here the Legendre pseudospectral method is employed, with $x_0=-1$, $x_N=1$; while  $x_j (j= \nolinebreak 1,\ldots,N-1)$ are to be obtained 
from roots of first derivative of Legendre polynomial $P_N(x)$ with respect to $x$ $(P'_N(x_j) = 0).$
The $g_j(x)$ in Eq.~(3) are called cardinal functions satisfying the relation, $g_j(x_{j'}) = \delta_{j'j}$. At this stage, the semi-infinite 
domain $r \in [0, \infty]$ is mapped onto the finite domain $x \in [-1,1]$ through a transformation $r=r(x)$. Next, a nonlinear algebraic mapping 
of the following form is introduced, for convenience, 
\begin{equation}
r=r(x)=L\ \ \frac{1+x}{1-x+\alpha},
\end{equation}
where L and $\alpha=2L/r_{max}$ are two adjustable mapping parameters. Then, introduction of a transformation of the type
$\psi(r(x))=\sqrt{r'(x)} f(x), $ followed by a symmetrization procedure leads to a \emph {symmetric} matrix eigenvalue problem.
This is easily solved by standard available routines (NAG libraries, for example) offering quite accurate eigenvalues and eigenfunctions. 

\section{Results and Discussion}
%\begin{enumerate}
\subsection {Confined H atom} 

\begingroup
\squeezetable
\begin{table}
\caption {\label{tab:table1}Energies (a.u.) of CHA for low-lying states. PR implies Present Result.} 
\begin{ruledtabular}
\begin{tabular}{lll|lll}
$r_c$  & E$_{1s}$ (PR)   & E$_{1s}$ (Literature) &  $r_c$        &       E$_{2s}$ (PR)  &  E$_{2s}$ (Literature)   \\
\hline
0.1    & 468.9930386595      & 468.9930\footnotemark[1],468.9930386593\footnotemark[2],468.99313\footnotemark[3]   &  
0.1    & 1942.720354554      & 1942.720\footnotemark[1]                                                          \\ 
0.2    & 111.0698588367      & 111.0698588368\footnotemark[2],111.07107\footnotemark[3]                             &
0.2    & 477.8516723922      &                                                                                   \\   
0.5    & 14.74797003035      & 14.74797\footnotemark[1],14.74797003035\footnotemark[2]$^,$\footnotemark[4],14.74805\footnotemark[3]   &
0.5    & 72.67203919047      & 72.67204\footnotemark[1],72.67203919046\footnotemark[4]                           \\
1.0    & 2.373990866100      & 2.373991\footnotemark[1],2.373990866103\footnotemark[2]$^,$\footnotemark[4],      & 
1.0    & 16.57025609346      & 16.57026\footnotemark[1],16.57025609346\footnotemark[4],                          \\
       &                     & 2.37399\footnotemark[3],2.373990866\footnotemark[5]                                                       &
       &                     & 16.570256093\footnotemark[5]                                                      \\
1.5    & 0.437018065247      & 0.437018065247\footnotemark[4]                                                    &
2.0    & 3.327509156489      & 3.327509\footnotemark[1],3.32750915649\footnotemark[6]                            \\  
2.0    & $-$0.125000000002   & $-$0.1250000\footnotemark[1],$-$0.125000000000\footnotemark[2]$^,$\footnotemark[4], &
4.0    & 0.420235631712      & 0.4202356\footnotemark[1],0.420235631713\footnotemark[4],         \\
       &                     & $-$0.12500\footnotemark[3],$-$0.125000000\footnotemark[5],$-$0.12500000000\footnotemark[6]                    &
       &                     &                                                                   \\
3.0    & $-$0.423967287733   & $-$0.423967287733\footnotemark[2]$^,$\footnotemark[4]                              &
6.0    & 0.012725103091      & 0.01272510\footnotemark[1],0.012725103090\footnotemark[4]                          \\ 
4.0    & $-$0.483265302077   & $-$0.4832653\footnotemark[1],$-$0.483265302078\footnotemark[2]$^,$\footnotemark[4], &   
8.0    & $-$0.084738721360   & $-$0.08473872\footnotemark[1],$-$0.0847387213569\footnotemark[4],                  \\
       &                     & $-$0.48327\footnotemark[3]                                                         &
       &                     & $-$0.084738721\footnotemark[5],$-$0.08473872135\footnotemark[6]$^,$\footnotemark[7]  \\
5.0    & $-$0.496417006591   & $-$0.496417006591\footnotemark[4]                                                  &
10.0   & $-$0.112806210298   & $-$0.1128062\footnotemark[1],$-$0.112806210295\footnotemark[4],  \\
       &                     &                                                                                    &
       &                     & $-$0.11280621029\footnotemark[6],$-$0.112806210296\footnotemark[7]                  \\
6.0    & $-$0.499277286372   & $-$0.4992773\footnotemark[1],$-$0.499277286372\footnotemark[4],                    &
14.0   & $-$0.124015029434   & $-$0.124015029431\footnotemark[4],$-$0.124015029\footnotemark[5],                   \\
       &                     & $-$0.49928\footnotemark[3]                                                         &
       &                     & $-$0.12401502943\footnotemark[6],$-$0.124015029432\footnotemark[7]                  \\  
8.0    & $-$0.499975100445   & $-$0.4999751\footnotemark[1],$-$0.49997\footnotemark[3],$-$0.499975100445\footnotemark[4],                     &
20.0   & $-$0.124987114308   & $-$0.124987114312\footnotemark[4],$-$0.124987114\footnotemark[5],                   \\
       &                     & $-$0.499975100\footnotemark[5],$-$0.49997510044\footnotemark[6],                    &
       &                     & $-$0.124987114313\footnotemark[7]                                                   \\   
       &                     & $-$0.499975100446\footnotemark[7]                                                   &
       &                     &                                                                                     \\
10.0   & $-$0.499999263281   & $-$0.4999993\footnotemark[1],$-$0.50000\footnotemark[3],$-$0.499999263281\footnotemark[4],                     &
25.0   & $-$0.124999763707   & $-$0.1249998\footnotemark[1],$-$0.12499976370\footnotemark[6]                       \\
       &                     & $-$0.49999926328\footnotemark[6],$-$0.499999263282\footnotemark[7]                  &
       &                     &                                                                                     \\
12.0   & $-$0.499999980159   & $-$0.49999998015\footnotemark[6],$-$0.499999980159\footnotemark[7]                  &
30.0   & $-$0.124999996469   & $-$0.12499999646\footnotemark[6]                                                    \\
20.0   & $-$0.500000000000   & $-$0.499999999999\footnotemark[4],$-$0.500000000\footnotemark[5]                    &
40.0   & $-$0.124999999998   & $-$0.124999999999\footnotemark[7]                                                   \\   
\end{tabular}
\end{ruledtabular}
\begin{tabbing}
$^{\mathrm{a}}$Ref.~\cite{goldman92}. \hspace{15pt}  \=
$^{\mathrm{b}}$Ref.~\cite{ciftci09}. \hspace{15pt}  \=
$^{\mathrm{c}}$Ref.~\cite{aquino14}. \hspace{15pt}  \=
$^{\mathrm{d}}$Ref.~\cite{aquino07}. \hspace{15pt}  \=
$^{\mathrm{e}}$Ref.~\cite{burrows06}. \hspace{15pt}  \=
$^{\mathrm{f}}$Ref.~\cite{aquino95}. \hspace{15pt}  \=
$^{\mathrm{g}}$Ref.~\cite{laughlin02}. 
\end{tabbing}
\end{table}
\endgroup

At first, Table I gives energies of H atom at the center of an inert impenetrable cavity for two lowest-lying $s$ states, \emph{viz.}, 
$1s$ and $2s$. It is worth mentioning at the outset that, henceforth all quantities are given in atomic unit, unless mentioned otherwise. 
We have carefully selected 14 $r_c$ to cover \emph{small, intermediate and large} range of confinement. As mentioned previously, 
a host of results for such low states exists. Best six of them are chosen as reference for comparison. For the entire range
of radius, energies were computed to seven-figure accuracy in \cite{goldman92}. Exact energies, expressed through Kummer $M-$functions 
\cite{burrows06}, are also available for these states. For all values of radii, present energies for both states completely agree with the
quoted values, for up to the precision they are presented in \cite{burrows06}. Some other very accurate results are also found, e.g.,
the series method \cite{aquino07}, asymptotic iteration method \cite{ciftci09}. Former results exist for both states, while same for the
latter offers only $2s$ states. For all instances, our energies are seen to either agree completely with these, or differ only in the last place 
of decimal quoted. Reasonably accurate eigenvalues of these states are also reported for intermediate to large $r_c$ values in \cite{aquino95}
and for $r_c \geq 8$ in \cite{laughlin02}. Qualitatively correct energies for ground state were obtained from some simple variational 
wave functions \cite{varshni99}, and a variational method with generalized Hylleraas basis set as well as a perturbative approach
using exact solution of confined free particles as unperturbed wave function \cite{aquino14}. As seen, energy levels are raised relative to the 
free-atom values. Obviously, as $r_c$ tends to 
infinity, eigenvalues monotonically approach the corresponding values of free H atom. With an increase in quantum number $n$, the required
$r_c$ values to attend the energy of unconfined H atom, increases. One also notices that the extent by which a CHA level is raised relative
to the free H atom, tends to increase as $r_c$ decreases.

\begingroup
\squeezetable
\begin{table}
\caption {\label{tab:table2}Energies (a.u.) of CHA for low-lying states. PR implies Present Result.} 
\begin{ruledtabular}
\begin{tabular}{lll|lll}
$r_c$  & E$_{2p}$ (PR)   & E$_{2p}$ (Literature) &  $r_c$        &       E$_{3p}$ (PR)  &  E$_{3p}$ (Literature)   \\
\hline
0.1    & 991.0075894412    & 991.0076\footnotemark[1]                                                                                 &
0.1    & 2960.462302278    & 2960.462\footnotemark[1]                                                                                 \\
0.2    & 243.1093166600    &                                                                                                          &
0.2    & 734.2292278041    &                                                                                                          \\
0.5    & 36.65887588018    & 36.65888\footnotemark[1],36.703\footnotemark[2],36.65887588018\footnotemark[3]                            &
0.5    & 114.6435525192    & 114.6436\footnotemark[1],114.6435525192\footnotemark[3]                                                   \\
1.0    & 8.223138316165    & 8.223138\footnotemark[1],8.233\footnotemark[2],8.223138316160\footnotemark[3],8.232\footnotemark[4],      &
1.0    & 27.47399530254    & 27.47400\footnotemark[1],27.47399530253\footnotemark[3],                                                  \\
       &                   & 8.223138316\footnotemark[5],8.223138316160\footnotemark[6]                                                &
       &                   & 27.473995303\footnotemark[5]                                                                              \\
2.0    & 1.576018785601    & 1.576019\footnotemark[1],1.57775\footnotemark[2],1.576018785606\footnotemark[3]$^,$\footnotemark[6],      &
2.0    & 6.269002791978    & 6.269003\footnotemark[1],6.269002791986\footnotemark[3],                                                  \\
       &                   & 1.57735\footnotemark[4],1.576018786\footnotemark[5],1.57601878560\footnotemark[7]                         &
       &                   & 6.269002792\footnotemark[5]                                                                               \\
4.0    & 0.143527083713    & 0.1435271\footnotemark[1],0.14366\footnotemark[2],0.143527083713\footnotemark[3]$^,$\footnotemark[6],     &
5.0    & 0.707718415829    & 0.707718415822\footnotemark[3]                                                                            \\
       &                   & 0.14359\footnotemark[4],0.14352708371\footnotemark[7]                                                    &
       &                   &                                                                                                           \\
6.0    & $-$0.055555555557 & $-$0.05555556\footnotemark[1],$-$0.055555\footnotemark[2]$^,$\footnotemark[4],$-$0.055555555555\footnotemark[3] &
10.0   & 0.049190760574    & 0.04919076\footnotemark[1],0.049190760586\footnotemark[3]                                                 \\
8.0    & $-$0.104450066408 & $-$0.1044501\footnotemark[1],$-$0.10441\footnotemark[2],
                             $-$0.104450066406\footnotemark[3]$^,$\footnotemark[6]$^,$\footnotemark[8],                                &
14.0   & $-$0.027268482516 & $-$0.027268482486\footnotemark[3],                                                                        \\   
       &                   & $-$0.104450066\footnotemark[5],$-$0.10445006640\footnotemark[7]                                           &
       &                   & $-$0.027268482\footnotemark[5]                                                                            \\
10.0   & $-$0.118859544856 & $-$0.1188595\footnotemark[1],$-$0.118859544853\footnotemark[3],                                           &
20.0   & $-$0.051611419756 & $-$0.051611419761\footnotemark[3],                                                                        \\
       &                   & $-$0.11885954485\footnotemark[7],$-$0.118859544854\footnotemark[8]                                        &
       &                   & $-$0.051611420\footnotemark[5]                                                                            \\
14.0   & $-$0.124540597991 & $-$0.124540597990\footnotemark[3]$^,$\footnotemark[8],$-$124540598\footnotemark[5],$-$.12454059799\footnotemark[7]  &
25.0   & $-$0.054909464520 & $-$0.05490946\footnotemark[1]                                                                             \\
20.0   & $-$0.124994606646 & $-$124994606647\footnotemark[3]$^,$\footnotemark[8],$-$0.124994607\footnotemark[5],$-$0.12499460664\footnotemark[7] &
30.0   & $-$0.055471281464 &                                                                                                           \\
25.0   & $-$0.124999906046 & $-$0.1249999\footnotemark[1],$-$0.12499990604\footnotemark[7]                                              &
40.0   & $-$0.055554769957 &                                                                                                            \\
30.0   & $-$0.124999998641 & $-$0.12499999864\footnotemark[7]                                                                           &
50.0   & $-$0.055555551158 & $-$0.05555555\footnotemark[1]                                                                              \\
40.0   & $-$0.124999999999 &                                                                                                            &
55.0   & $-$0.055555555273 &                                                                                                            \\ 
\end{tabular}
\end{ruledtabular}
\begin{tabbing}
$^{\mathrm{a}}$Ref.~\cite{goldman92}. \hspace{15pt}  \=
$^{\mathrm{b}}$Ref.~\cite{filho02}. \hspace{15pt}  \=
$^{\mathrm{c}}$Ref.~\cite{aquino07}. \hspace{15pt}  \=
$^{\mathrm{d}}$Ref.~\cite{varshni97}. \hspace{15pt}  \=
$^{\mathrm{e}}$Ref.~\cite{burrows06}. \hspace{15pt}  \=
$^{\mathrm{f}}$Ref.~\cite{ciftci09}. \hspace{15pt}  \=
$^{\mathrm{g}}$Ref.~\cite{aquino95}. \hspace{15pt}  \=
$^{\mathrm{h}}$Ref.~\cite{laughlin02}. \hspace{15pt}  \=
\end{tabbing}
\end{table}
\endgroup

Next, Table II offers energies of two low-lying excited states having $\ell=1$, namely, $2p$ and $3p$, of a CHA for varying radii (at 14 selected)
of the enclosure. Once again, a decent number of theoretical results are available for these states, especially in the intermediate $r_c$; some of 
those are duly quoted here for comparison. For the asymptotically small radius ($r_c < 0.5$), only one result could be found; present
energies are clearly much better than the reference values \cite{goldman92}. On the other hand, the intermediate-$r_c$ results for $2p$ state are 
seen to match quite nicely with the algebraic solution of \cite{laughlin02}. As in the previous table, current eigenvalues are very much competitive 
to those of \cite{aquino07,burrows06,ciftci09}; in several occasions completely reproducing the latter ones. For $2p$ state, some 
intermediate-$r_c$ results were reported through a super-symmetric variational method \cite{filho02} as well as a variational 
method \cite{varshni97}, offering qualitatively good energies. Free-atom energies are regained back for sufficiently large box size $r_c$; 
large $n$ requires large $r_c$. Other general conclusions of Table I remain valid here also.  

\begingroup
\squeezetable
\begin{table}
\caption {\label{tab:table3}Energies (a.u.) of CHA for some high-lying states. PR implies Present Result.} 
\begin{ruledtabular}
\begin{tabular}{ccclccl}
State & $r_c$ & Energy (PR)  & Energy (Reference)        &   $r_c$      &  Energy (PR)   &   Energy (Reference) \\ 
\hline
$3s$  & 0.1  & 4406.1216518      & 4406.122\footnotemark[1]             &
        1    & 40.863124601      & 40.86312\footnotemark[1],40.863124601\footnotemark[2]$^,$\footnotemark[3]           \\
      & 5    & 1.0532206154      & 1.0532206155\footnotemark[3]         &
        15   & $-$0.0268748755   &                                      \\
      & 25   & $-$0.0545924509   & $-$0.05459245\footnotemark[1]        &
        50   & $-$0.0555555478   & $-$0.05555555\footnotemark[1]        \\
$4s$  & 0.1  & 7857.6291849      & 7857.629\footnotemark[1]             &
        1    & 75.130493060      & 75.13049\footnotemark[1],75.130493061\footnotemark[2]$^,$\footnotemark[3]          \\
      & 5    & 2.3823251868      & 2.3823251868\footnotemark[3]         &
        25   & $-$0.0132027435   & $-$0.01320274\footnotemark[1]        \\
      & 40   & $-$0.0305518195   &                                      &
        50   & $-$0.0312043375   & $-$0.03120434\footnotemark[1]        \\
      & 60   & $-$0.0312481497   &                                      &
        80   & $-$0.0312499987   &                                      \\
$5s$  & 0.1  & 12296.731659      & 12296.73\footnotemark[1]             &
        1    & 119.32706249      & 119.3271\footnotemark[1],119.327062496\footnotemark[2]$^,$\footnotemark[3]          \\
      & 10   & 0.8263888778      & 0.8263889\footnotemark[1],0.8263888778\footnotemark[3]            &
        20   & 0.1128777394      & 0.112877739\footnotemark[2],0.1128777394\footnotemark[3]          \\
      & 40   & $-$0.0110593683   &                                      &
        60   & $-$0.0195964955   &                                      \\
      & 80   & $-$0.0199943334   &                                      &
        100  & $-$0.0199999715   &                                      \\
$4p$  & 0.1  & 5918.1828889      & 5918.183\footnotemark[1]             &
        1    & 56.758033888      & 56.75803\footnotemark[1],56.758033888\footnotemark[2]             \\
      & 5    & 1.8304233586      &                                      &
        25   & $-$0.0165034620   & $-$0.01650346\footnotemark[1]        \\
      & 40   & $-$0.0307098946   &                                      &
        50   & $-$0.0312164983   & $-$0.03121650\footnotemark[1]        \\
      & 60   & $-$0.0312486974   &                                      &
        80   & $-$0.0312499991   &                                      \\
$5p$  & 0.1  & 9863.6047594      & 9863.605\footnotemark[1]             &
        1    & 95.991853334      & 95.99185\footnotemark[1],95.991853335\footnotemark[2]             \\
      & 10   & 0.6883703331      & 0.6883703\footnotemark[1]            &
        20   & 0.0951697270      &                                      \\
      & 40   & $-$0.0122109669   &                                      &
        60   & $-$0.0196650907   &                                      \\
      & 80   & $-$0.0199955819   &                                      &
        100  & $-$0.0199999786   &                                      \\
$3d$  & 0.1  & 1644.5299223      & 1644.530\footnotemark[1]             &
        1    & 14.967464086      & 14.96746\footnotemark[1],14.9895\footnotemark[4],14.979\footnotemark[5]        \\
      & 5    & 0.3291171429      & 0.329425\footnotemark[4],0.329365\footnotemark[5]                     &
        15   & $-$0.0466425817   & $-$0.046635\footnotemark[5]                                  \\
      & 25   & $-$0.0553214524   & $-$0.05532145\footnotemark[1],$-$0.05532145239\footnotemark[6]        &
        50   & $-$0.0555555544   & $-$0.05555555\footnotemark[1],$-$0.0555555544\footnotemark[6]        \\
$4d$  & 0.1  & 4115.5826320      & 4115.583\footnotemark[1]             &
        1    & 39.315319855      & 39.31532\footnotemark[1]             \\
      & 5    & 1.2396510218      &                                      &
        25   & $-$0.0218552724   & $-$0.02185527\footnotemark[1]        \\
      & 40   & $-$0.0309495183   &                                      &
        50   & $-$0.0312333327   & $-$0.03123333\footnotemark[1]        \\
      & 60   & $-$0.0312494032   &                                      &
        80   & $-$0.0312499996   &                                      \\
$5d$  & 0.1  & 7569.5425196      & 7569.543\footnotemark[1]             &
        1    & 73.601919340      & 73.60192\footnotemark[1]             \\
      & 10   & 0.5213480960      & 0.5213481\footnotemark[1]            &
        20   & 0.0682442940      &                                      \\
      & 40   & $-$0.0142130246   &                                      &
        60   & $-$0.0197771075   &                                      \\
      & 80   & $-$0.0199974140   &                                      &
        100  & $-$0.0199999884   &                                      \\
$4f$  & 0.1  & 2426.3955489      & 2426.396\footnotemark[1]             &
        1    & 22.895825482      & 22.89583\footnotemark[1]             \\
      & 5    & 0.6694519803      &                                      &
        25   & $-$0.0274353350   & $-$0.02743534\footnotemark[1]        \\
      & 40   & $-$0.0311568571   &                                      &
        50   & $-$0.0312456821   & $-$0.03124568\footnotemark[1]        \\
      & 60   & $-$0.0312498630   &                                      &
        80   & $-$0.0312499999   &                                      \\
$5f$  & 0.1  & 5407.2220240      & 5407.222\footnotemark[1]             &
        1    & 52.395102963      & 52.39510\footnotemark[1]             \\
      & 10   & 0.3525841702      & 0.3525842\footnotemark[1]            &
        20   & 0.0384456748      &                                      \\
      & 40   & $-$0.0165731807   &                                      &
        60   & $-$0.0198918442   &                                      \\
      & 80   & $-$0.0199989716   &                                      &
        100  & $-$0.0199999958   &                                      \\
$5g$  & 0.1  & 3333.3040034      & 3333.304\footnotemark[1]             &
        1    & 32.034089112      & 32.03409\footnotemark[1]             \\
      & 10   & 0.1883418745      & 0.1883419\footnotemark[1]            &
        20   & 0.0090019053      &                                      \\
      & 40   & $-$0.0187098818   &                                      &
        60   & $-$0.0199708957   &                                      \\
      & 80   & $-$0.0199997891   &                                      &
        100  & $-$0.0199999992   &                                      \\
\end{tabular}
\end{ruledtabular}
\begin{tabbing}
$^{\mathrm{a}}$Ref.~\cite{goldman92}. \hspace{35pt}  \=
$^{\mathrm{b}}$Ref.~\cite{burrows06}. \hspace{35pt}  \=
$^{\mathrm{c}}$Ref.~\cite{aquino07}.  \hspace{35pt}  \=
$^{\mathrm{d}}$Ref.~\cite{varshni97}. \hspace{35pt}  \=
$^{\mathrm{e}}$Ref.~\cite{filho02}.   \hspace{35pt}  \=
$^{\mathrm{f}}$Ref.~\cite{aquino95}.  
\end{tabbing}
\end{table}
\endgroup

Now Table III offers results on select 12 low- and moderately-high-lying excited states of CHA corresponding to 
$3 \leq n \leq 5; \ \ell \leq 4$, to 
establish the efficiency and usefulness of our present method in confined situations. In all cases, a wide region of confinement has been 
considered. Unlike the previous tables, in this case, reference results are rather scarce, which are quoted appropriately. For all these
states, some results are available from the numerical calculation of \cite{goldman92}, showing a decent agreement with ours. For $s$, $p$
states, best reference energies are those from \cite{aquino95} and \cite{burrows06}; gratifyingly, present eigenvalues are in excellent 
agreement with these. For $d$ series, best literature energies seem to be those reported in \cite{aquino95}; here also one notices
good matching between present and literature values. No reference values other than those in \cite{goldman92} could be found for the 
$\ell > 2$ series for comparison. As seen, in all cases, present results are significantly improved from the reference ones.
For $3d$ states of CHA, reasonably good energies are also reported in the variational \cite{varshni97} and super-symmetric \cite{filho02}
calculation. As can be seen, many of these states have not been reported earlier, especially for medium and large $r_c$. Given the success of 
this method for all the states reported before, we are confident that these are also equally accurate; they may constitute a useful 
reference for future investigations on such systems.

\begingroup
\squeezetable
\begin{table}
\caption {\label{tab:table4}Eigenvalues (in a.u.) of $n=8$ states of 3D CHA as function of $r_c$.} 
\begin{ruledtabular}
\begin{tabular}{cccccccc}
$\ell$  & $r_c=1$    & $r_c=25$        &    $r_c=50$      &  $r_c=75$       &   $r_c=100$     &  $r_c=150$     &  $r_c=175$  \\
\hline
0   &  311.32325639  & 0.328725861     & 0.04409216       &  0.00483160     & $-$0.00466283   & $-$0.0077439   & $-$0.0078087  \\
1   &  273.14306704  & 0.301937308     & 0.04108258       &  0.00416432     & $-$0.00484461   & $-$0.0077505   & $-$0.0078091  \\
2   &  235.92204657  & 0.264080555     & 0.03584294       &  0.00292353     & $-$0.00518755   & $-$0.0077624   & $-$0.0078098  \\
3   &  199.97204134  & 0.222512230     & 0.02927280       &  0.00125898     & $-$0.00565391   & $-$0.0077768   & $-$0.0078107  \\
4   &  165.32153876  & 0.180052301     & 0.02204074       & $-$0.00066615   & $-$0.00619372   & $-$0.0077909   & $-$0.0078115  \\
5   &  131.87629255  & 0.137914527     & 0.01458520       & $-$0.00269811   & $-$0.00674844   & $-$0.0078021   & $-$0.0078120  \\
6   &  99.330037127  & 0.096520395     & 0.00718594       & $-$0.00469286   & $-$0.00725165   & $-$0.0078089   & $-$0.0078123  \\
7   &  66.624595253  & 0.055270190     & $-$0.00001574    & $-$0.00649140   & $-$0.00762894   & $-$0.0078118   & $-$0.0078124  \\
\end{tabular}
\end{ruledtabular}
\end{table}
\endgroup

Once the low-lying states of Tables I, II and III are obtained, an extension is made for some high-lying excited states,
as a further illustration of feasibility and performance of the approach. As a representative set, all 8 states belonging to 
$\ell=0-7$ corresponding to $n=8$ of CHA are tabulated in Table IV. Higher states have been scarcely dealt in literature; thus 
no references exist. Eigenvalues for all these are given at seven selected $r_c$ values, namely, 1, 25, 50, 75, 100, 150, 175 a.u. to 
cover a broad region of confinement. Within a particular $n$, for a fixed $r_c$, energies are split such that, with increase in 
$\ell$, the latter decreases, so that the sub-level with \emph{largest} $\ell$ corresponds to \emph{lowest} energy. Obviously, as 
$r_c \rightarrow \infty$, all states regain the free-atom energies. A careful examination of the above tables reveals that, for a 
given $r_c$, the extent by which a particular level is raised with respect to its corresponding free state, is relatively less for
ground state compared to excited state; generally increasing with $n$ for a specific $\ell$. For example, the magnitude of the shift
in energy from its unconfined counterpart, $\Delta E_{n\ell}^{r_c}=E_{n\ell}^{r_c}-E_{n\ell}$, for $1s$ are: $1.4 \times 10^{-4}, 2.5 \times 10^{-5},
7.4 \times 10^{-7}, 2.0 \times 10^{-8}, 5.0 \times 10^{-10},$ for $r_c= 7, 8, 10, 12, 14$ respectively, while the same values for $2s$ 
are: $7.4 \times 10^{-2}, 4.0 \times 10^{-2}, 1.2 \times 10^{-2}, 3.6 \times 10^{-3}, 9.8 \times 10^{-4}$.  
These differences decrease in an exponential fashion with increasing $r_c$ and then finally vanish for a wall placed at sufficiently large 
distance; which is consistent with the findings of \cite{laughlin02, shaqqor09}. It is hoped that, in future, these results would be helpful for 
calibration of other methods.

\begin{figure}
\centering
\begin{minipage}[t]{0.40\textwidth}\centering
\includegraphics[scale=0.38]{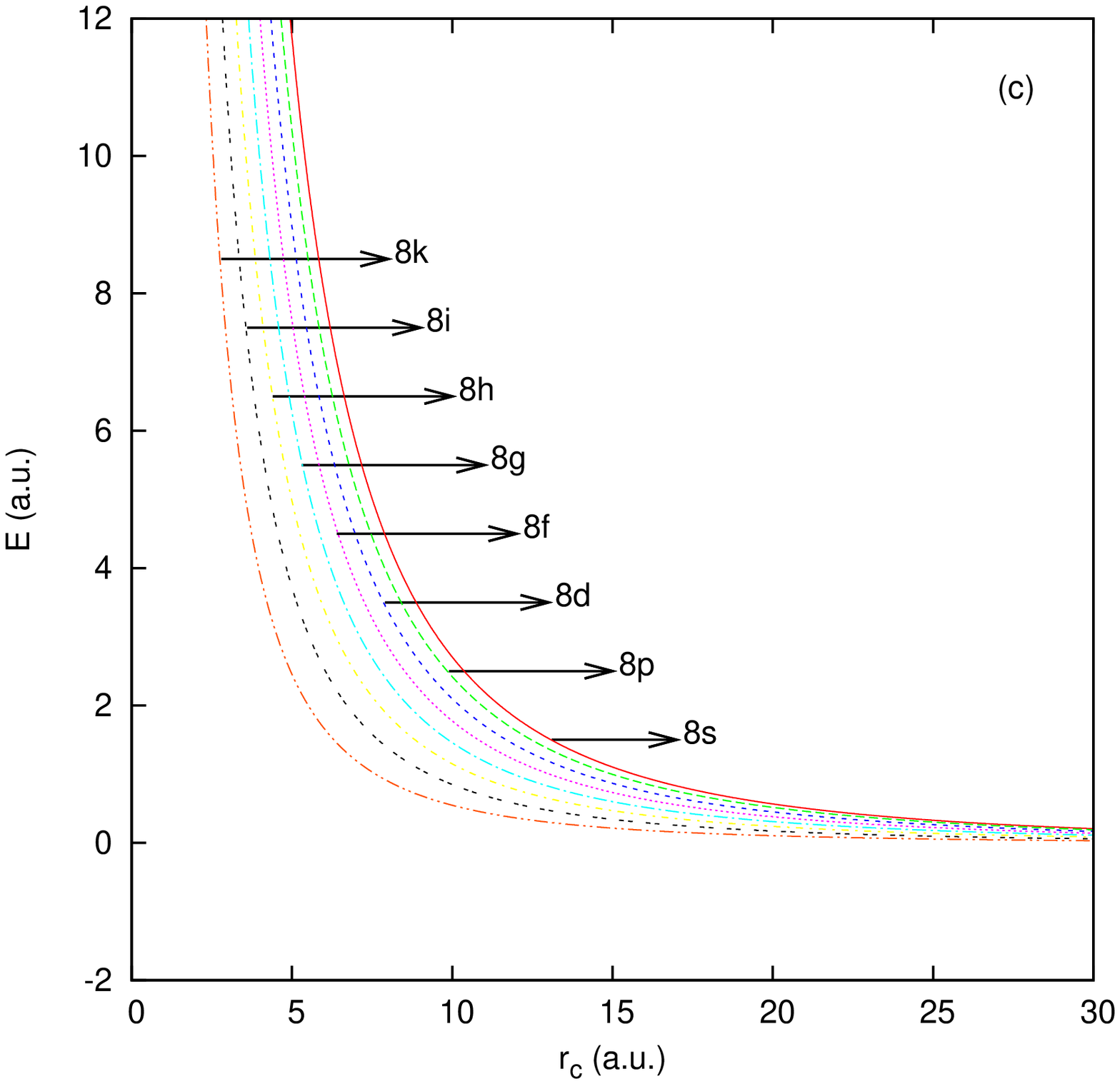}
\end{minipage}
\hspace{0.15in}
\begin{minipage}[t]{0.35\textwidth}\centering
\includegraphics[scale=0.38]{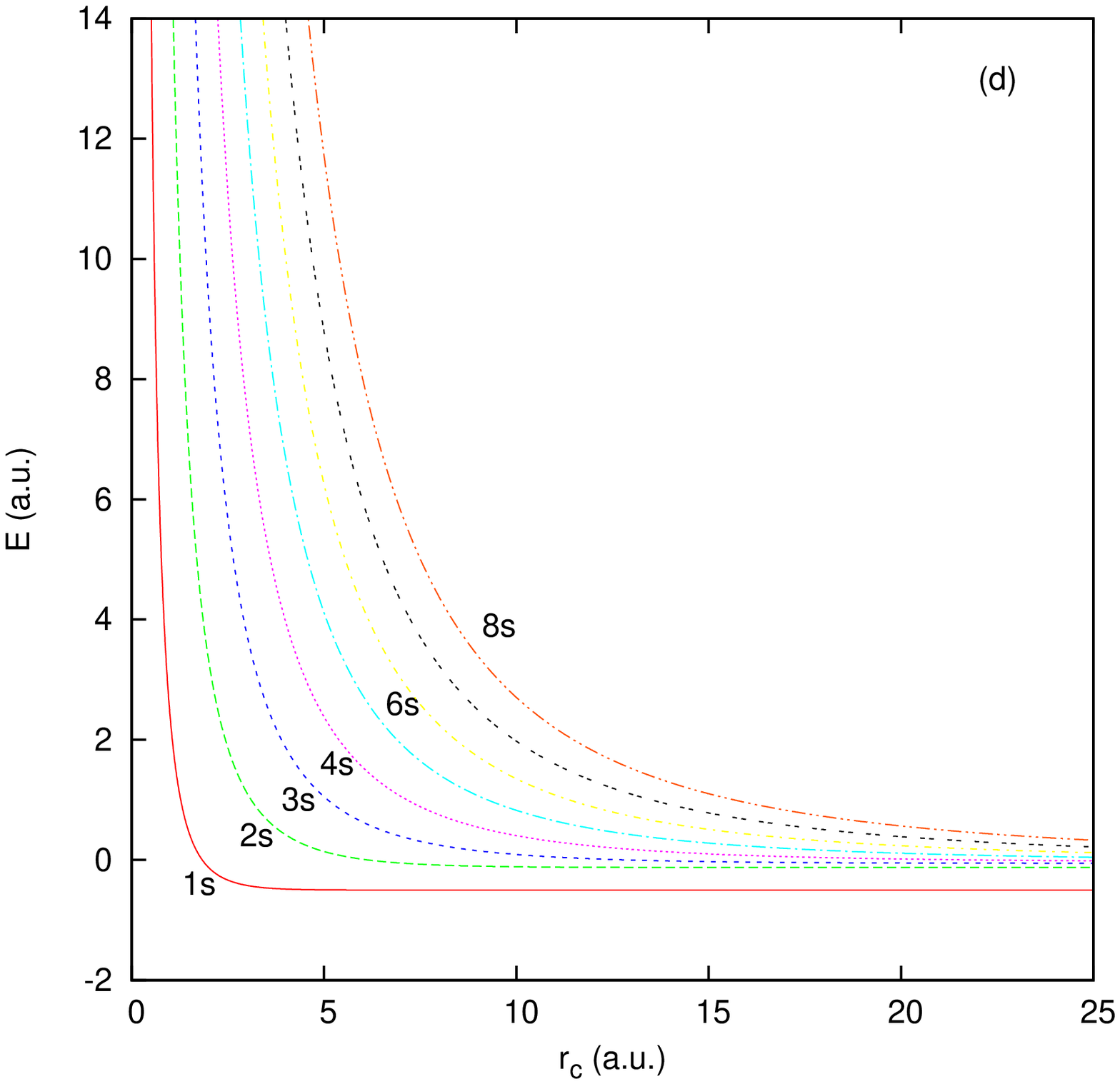}
\end{minipage}
\\[10pt]
\begin{minipage}[b]{0.40\textwidth}\centering
\includegraphics[scale=0.38]{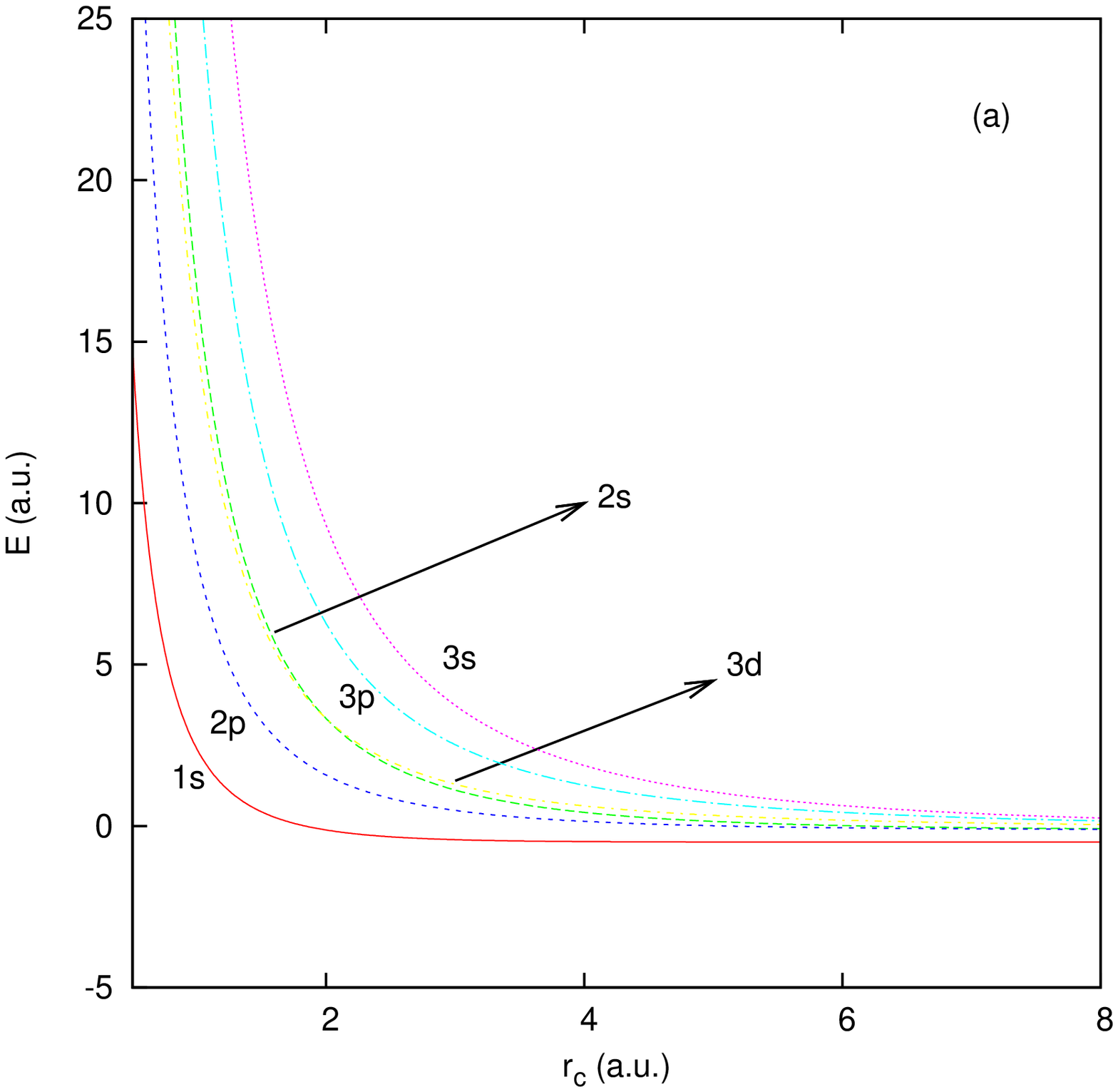}
\end{minipage}
\hspace{0.15in}
\begin{minipage}[b]{0.35\textwidth}\centering
\includegraphics[scale=0.38]{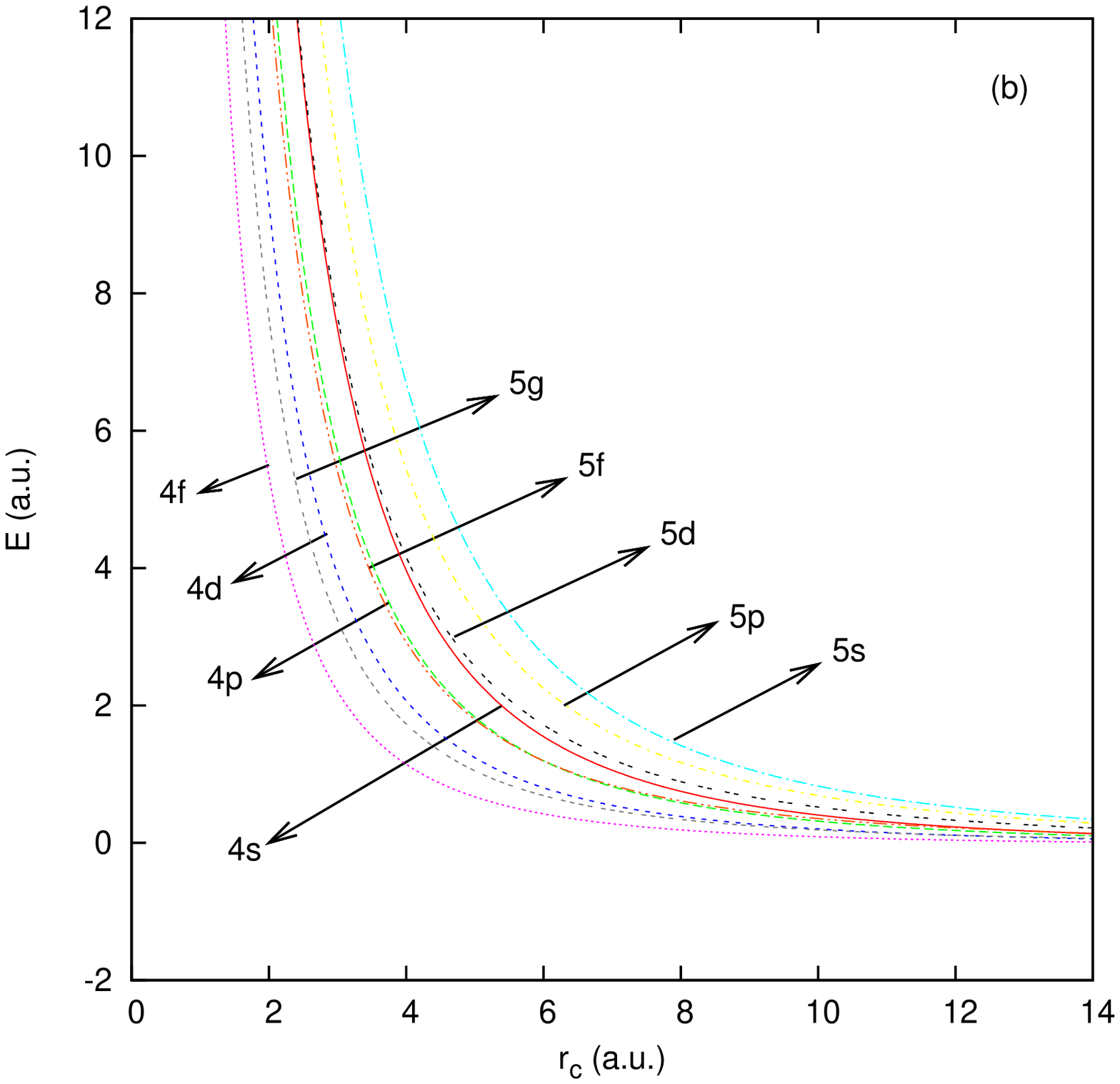}
\end{minipage}
\caption[optional]{Energy variations in CHA with $r_c$: (a) 6 states corresponding to $n=1,2,3$ (b) 9 states with $n=4,5$ 
(c) Eight $n=8$ states having $\ell=0-7$ (d) Eight $\ell=0$ states having $n=1-8.$}
\end{figure}

Above energy variations of CHA in Tables I--IV are graphically shown in Fig.~1 for medium to large-size cavity. Note that, ranges of 
$r_c$ and energy axes are different for different plot. Also in all the cases, confinement in very small-sized box is ignored as this gives rise 
to very high energy values making the plots difficult to visualize. Panels (a), (b) depict energies of all six ($1s, 2s, 2p, 3s, 3p, 3d$) states 
corresponding to $n=1-3$ and all nine states for $n=4,5$ respectively, with changes in $r_c$. It is well-known that the characteristic 
\emph{accidental} degeneracy of free H atom (a consequence of the central Coulombic field), is broken in presence of a hard impenetrable wall 
at \emph{all} finite radius $r_c$. This is due to the fact that confinement results in a violation of the requirement of the potential being 
purely Coulombic everywhere. The adjacent plots remain parallel to each other. General nature of the plots are consistent with those of a 
central potential under an isotropic confinement, i.e., very high energy at small $r_c$ followed by a sharp decline with an increase in $r_c$, 
finally attaining the value of that of respective free atom at a sufficiently large $r_c$ and remaining constant afterwards. In all cases, 
individual confined energy levels are raised relative to the unconfined H case. For a given state, the magnitude by which this raise occurs, 
increases as $r_c$ assumes progressively smaller values. This is promptly verified from the $\Delta_{n\ell}^{r_c}$ values, as defined earlier, 
for example, for $2p$ state, as in the sequence: $6.9 \times 10^{-2}, 6.1 \times 10^{-3}, 2.3 \times 10^{-4}, 5.4 \times 10^{-6}, 9.4 \times 10^{-8},
1.4 \times 10^{-9}$, corresponding to $r_c$ values of $6, 10, 15, 20, 25, 30$ respectively. Such a system is also known to exhibit 
\emph{simultaneous degeneracy}, whereby, for 
all $n \geq \ell+2$, a CHA state characterized by quantum numbers $(n,\ell)$ becomes degenerate to a $(n+1, \ell+2)$ state, exactly at 
$r_c=(\ell+1)(\ell+2)$. Thus it is seen that $2s$ and $3d$ states are degenerate at $r_c=2$. Some other similar pairs are $(4s,5d)$ at $r_c=2$, as 
well as $(3s, 4d)$, $(3p,4f)$, $(4p, 5f)$, all at $r_c =6$. Panel (c) displays all the eight states (having $\ell=0-7$) corresponding to $n=8$ 
as in Table IV. Within a given $n$, the sub-$\ell$ levels do not cross each other; all the plots remain well-separated at small $r_c$ and 
gradually reaches the free-atom value at large $r_c$. As $r_c$ decreases, higher-$\ell$ states get relatively more stabilized such that, for 
a particular $r_c$, accidental degeneracy breaks down to make the highest-$\ell$ state lowest in energy and vice versa \cite{fowler84}. 
Therefore, for a particular $n$, one finds inequalities such as: $E_{2p} < E_{2s}$; $E_{3d} < E_{3p} < E_{3s}$; 
$E_{4f} < E_{4d} < E_{4p} < E_{4s}$, etc. Lastly in (d) is shown the plots for eight $s$-waves ($\ell=0$) having radial quantum numbers 
$n=1-8$. Here also, the plots remain well-separated and monotonically decreasing with increase in $r_c$. For a given $\ell$ and $r_c$, state 
with lowest $n$ remains lowest in energy and vice versa. Similar pattern has been found for other $\ell$ values and omitted therefore. 
As expected, as the cavity size becomes smaller, many complex energy splitting is observed, especially with higher $n, \ell$ quantum numbers. 
We have found the energy orderings for CHA in the limit of $r_c \rightarrow 0$ as, 
\[ 1s,2p,3d,2s,4f,3p,5g,4d,6h,3s,5f,7i,4p,8k,6g,5d,4s,9l,7h,6f,10m,5p,8i, \cdots \]
It is noticed that as one goes to higher levels, there is significant intermixing between levels belonging to different $n$ values. This arises
presumably due to the fact that as confining radius decreases, there is a lot of crossover between levels of different $n,\ell$ values. 

\begingroup
\squeezetable
\begin{table}
\caption {\label{tab:table5} Estimated critical cage radius $r_c^c$ (a.u.) of CHA. All states having $n=1-10$ are given.} 
\begin{ruledtabular}
\begin{tabular}{l|cccccccccc}
$n$    & \multicolumn{10}{c}{State}  \\
\cline{2-11} 
     &   $s$   &  $p$       &   $d$  &   $f$   &   $g$  &         $h$  &  $i$  &   $k$           &   $l$        &   $m$               \\
\hline 
 1   &  1.8352463302                 &         &                  &    &       &     &           &      &       &                     \\ 
     &  1.8352463302\footnotemark[1] &        &                   &    &       &     &           &      &       &                     \\ 
 2   &  6.152307040                  &  5.0883082272                   &       &     &      &    &      &       &           &         \\ 
     &  6.152307040\footnotemark[1]  &  5.0883082272\footnotemark[1]   &       &     &      &    &      &       &           &         \\ 
 3   &  12.93743173                  &  11.90969656                    &  9.6173660416  &   &    &      &       &           &    &    \\ 
     &  12.93743173\footnotemark[1]  &  11.90969656\footnotemark[1]    &  9.6174\footnotemark[2] &  &   &       &   &       &    &    \\ 
 4   &  22.19009585                  &  21.1744312                     & 19.03014422             & 15.36345002  &   & &     &  & &    \\ 
     &  22.19009585\footnotemark[1]  &  21.1744312\footnotemark[1]     & 19.030\footnotemark[2]  & 15.363\footnotemark[2]   &  & &   &   &   &   \\ 
 5   &  33.9102067                   &  32.900106                      & 30.8119332              & 27.4587506               
     &  22.2921676                   &                                 &                         &                          &    &               \\ 
     &  33.9102067\footnotemark[1]   &  32.900106\footnotemark[1]      & 30.812\footnotemark[2]  & 27.459\footnotemark[2]   
     &  22.292\footnotemark[2]       &                                 &                         &                          &    &               \\ 
 6   &  48.097738                    &  47.090674                      & 45.030686               & 41.80445             
     &  37.15745                     &  30.380418                      &                         &                          &    &               \\ 
     &  48.097738\footnotemark[1]    &  47.090674\footnotemark[1]      & 45.031\footnotemark[2]  & 41.805\footnotemark[2]   
     &  37.157\footnotemark[2]       &  30.380\footnotemark[2]         &                         &                          &    &               \\ 
 7   &  64.752680                    &  63.747462                      & 61.703830               & 58.54454
     &  54.11667                     &  48.09827                       & 39.61139                &                          &    &               \\ 
     &  64.753\footnotemark[2]       &  63.747459\footnotemark[1]      & 61.704\footnotemark[2]  & 58.545\footnotemark[2]   
     &  54.117\footnotemark[2]       &  48.098\footnotemark[2]         & 39.611\footnotemark[2]  &                          &    &               \\ 
 8   &  83.874996                    &  82.87098                       & 80.83777                & 77.71881               
     &  73.41012                     &  67.72065                       & 60.2595                 & 49.9721                  &    &               \\ 
     &  83.875\footnotemark[2]       &  82.871\footnotemark[2]         & 80.838\footnotemark[2]  & 77.719\footnotemark[2]   
     &  73.410\footnotemark[2]       &  67.721\footnotemark[2]         & 60.260\footnotemark[2]  & 49.972\footnotemark[2]   &    &               \\ 
 9   &  105.466510                   &  104.46246                      & 102.43536               & 99.3425  
     &  95.1077                      &  89.6012                        & 82.5940                 & 73.6240                  & 61.452   &         \\ 
     &  105.46\footnotemark[2]       &  104.46\footnotemark[2]         & 102.44\footnotemark[2]  & 99.343\footnotemark[2]   
     &  95.108\footnotemark[2]       &  89.601\footnotemark[2]         & 82.594\footnotemark[2]  & 73.624\footnotemark[2]   & 61.452\footnotemark[2] & \\ 
 10  &  129.49108                    &  128.5179                       & 126.5288                & 123.4622  
     &  119.2614                     &  113.8539                       & 107.0970                & 98.718                   & 88.177   & 74.044   \\ 
     &  129.52\footnotemark[2]       &  128.52\footnotemark[2]         & 126.50\footnotemark[2]  & 123.42\footnotemark[2] 
     &  119.24\footnotemark[2]       &  113.85\footnotemark[2]         & 107.10\footnotemark[2]  & 98.718\footnotemark[2]   & 88.178\footnotemark[2] 
     &  74.045\footnotemark[2]       \\ 
\end{tabular}
\end{ruledtabular}
\begin{tabbing}
$^{\mathrm{a}}$ref.~\cite{ciftci09}. \hspace{35pt}  \=
$^{\mathrm{b}}$ref.~\cite{varshni98}. 
\end{tabbing}
\end{table}
\endgroup

Now the attention is turned to \emph{zero-energy} case, i.e., estimation of the minimum cavity radius that can accommodate a bound state in 
a CHA. As apparent from our above discussion, binding energy of a CHA gradually diminishes with reduction in the size of cavity, rendering all
states to have positive energy at sufficiently small $r_c$. Thus it is of importance to find out the cavity radius at which the binding 
energy becomes zero, the so-called \emph{critical cage radius}, $r_c^c$. For example, these have relevance in the study of partition function 
of atomic H as well as in the ionization of ground and excited state. Table V reports our calculated critical values for all 55 eigenstates 
of CHA, starting from ground state $1s$ to $10m$. Its first calculation for ground state of CHA produced a value of 1.835 a.u., which was
reported as early as in 1938 in the work of \cite{sommerfeld38}, through the zeros of Bessel's function of first kind of order $p$, $J_p(z)$. 
Thereafter, a slightly better value (1.8354) was presented in \cite{dingle53}. Some authors \cite{degroot46} also showed the ionization cage 
radii to be proportional to the zeros of Bessel functions. However, the first systematic investigation on all states up to and belonging to 
$n=10$ were undertaken by \cite{varshni98} through a variational calculation, which are duly quoted here for comparison. There, five 
significant-figure accurate results were given; current GPS results are considerably improved, especially for lower states. Results for first 
6 states corresponding to $\ell=0,1$ show \emph{complete} agreement with the accurate asymptotic iteration result \cite{ciftci09} for all but 
$7p$ state. First five states of $\ell=0,1$ are also available from precise calculations of \cite{burrows06}, which again corroborate our 
present critical radii values. Estimation of these become progressively more difficult for higher $n,\ell$ quantum numbers, due to complex 
mixing amongst states. In general, for a given $n$, critical radius tends to increase with $\ell$, while for a fixed $\ell$, the same 
decreases as $n$ increases. Furthermore, the disappearance of degeneracy in energy levels with respect to $\ell$ quantum number for a given $n$ 
in CHA is reminiscent to that of the effect of \emph{screening} on energy levels in a Coulomb potential \cite{roy05}, e.g., a Hulth\'en or 
Yukawa potential. It is well-known that in case of a \emph{screened} Coulomb potential, bound states exist only for certain values of screening
parameter below a threshold limit; if the parameter goes beyond this critical value, the state becomes unbound. Analogously, for H atom, under 
the influence of spherical confinement, a level becomes unbound if the confining radius remains below the critical cage radius, and bound otherwise.

\begingroup
\squeezetable
\begin{table}
\caption {\label{tab:table6} Selected expectation values (a.u.), for some low-lying states in CHA.} 
\begin{ruledtabular}
\begin{tabular}{ll|llll}
State  & $r_c$  &  $\langle r^{-2} \rangle $ & $\langle r^{-1} \rangle $ & $\langle r \rangle$ & $ \langle r^2  \rangle$ \\
\hline 
$1s$   &  0.5   &   40.6912615553   &  5.11404400581                  &  0.242490864909   &  0.067128353708                   \\
       &        &                   &  5.11404400581\footnotemark[1]  &                   &  0.067128353708\footnotemark[1]   \\
$3d$   &        &   11.5924623961   &  3.28244199632                  &  0.322576865990   &  0.108950015348                   \\
$1s$   &  2.0   &   4.10532919705   &  1.53516170643                  &  0.859353174267                   &  0.874825394135     \\
       &        &                   &  1.53516170643\footnotemark[1]  &  0.859353174266\footnotemark[1]   &  0.874825394134\footnotemark[1]     \\
$3d$   &        &   0.749825224105  &  0.832951857503                 &  1.27525204948    &  1.70676316887                    \\
       &        &                   &  0.832952\footnotemark[2]       &  1.275252\footnotemark[2]    &  1.706763\footnotemark[2]   \\
$1s$   & 10.0   &   1.99993975471   &  1.00001169282                  &  1.49993637877                    &  2.99945950887      \\
       &        &                   &  1.00001169282\footnotemark[1]  &  1.49993637877\footnotemark[1]    &  2.999459508865\footnotemark[1]     \\
$3d$   &        &   0.038607686219  &  0.185997997884                 &  5.84705927624    &  36.5172404570                    \\
\end{tabular}
\end{ruledtabular}
\begin{tabbing}
$^{\mathrm{a}}$Ref.~\cite{aquino07}. \hspace{35pt}  \=
$^{\mathrm{b}}$Ref.~\cite{montgomery01}. 
\end{tabbing}
\end{table}
\endgroup

As a further verification on the accuracy and faithfulness of our calculation, Table VI, additionally presents selected radial expectation
values, \emph{viz.}, $\langle r^{-2} \rangle $, $\langle r^{-1} \rangle $, $\langle r \rangle$, and $ \langle r^2  \rangle$ of CHA. For this,
$1s$ and $3d$ are chosen as representative states. The density moments have been reported earlier by many researchers; here the two best 
results are quoted. The overall qualitative agreement between present result and reference is quite good, again confirming the correctness 
and accuracy in our wave functions. Like the energy eigenvalues, position expectation values also monotonically approach the corresponding values of 
free H atom as the box radius tends to infinity. For some of them, no results could be found for comparison. 

\subsection{Confined Hulth\'{e}n Potential}
As an attempt to assess and extend the domain of applicability of our scheme to other Coulombic systems, at this stage, the focus is shifted
to the case of spherical confinement for the familiar short-range Hulth\'{e}n potential. It may be noted that, while 
many high-quality results are available for confined H atom, same for other Coulombic systems are rather scarce. Some notable confinement
works along this direction include (a) Hulth\'en potential \cite{sinha00,sinha03,filho03,varshni04,tian06} (b) Coulomb plus harmonic oscillator 
\cite{alberg01,hall11} (c) Morse potential \cite{silva10} (d) Lennard-Jones potential \cite{silva10a}, etc. Since maximum work has been 
done on (a), it is selected here to facilitate easy comparison. This potential shows Coulomb-like behavior for small $r$ and decays 
monotonically exponentially to zero for large $r$. However, interestingly, due to the presence of a screening parameter, it supports only a 
\emph{limited} number of bound states (unlike the Coulomb potential which possesses infinite number of states) for up to certain values of the 
parameter below a threshold limit. Furthermore, $\ell=0$ states of the \emph{free} system offer analytical solutions. Note that, a considerable 
amount of works exist for the free system. For example, quite accurate energies have been reported by a number of researchers 
\cite{nunez93,stubbins93,gonul00,roy05a}, which are characterized by complex level crossings for higher quantum numbers. 

\begingroup
\squeezetable
\begin{table}
\caption {\label{tab:table7} Comparison of some low-lying states of confined Hulth\'{e}n potential with literature data, for $\delta=0.1$. 
PR implies Present Result.} 
\begin{ruledtabular}
\begin{tabular}{l|ll|ll}
$r_c$  & E$_{1s}$(PR)      & E$_{1s}$(Ref.) &      E$_{2p}$(PR)   &  E$_{2p}$(Ref.)   \\
\hline
0.1    & 469.042997232     &                  &  991.057540215   &                   \\ 
0.5    & 14.7977679573     &                  &  36.7086314074   &                   \\
1.0    & 2.4236006207      &                  &  8.27265368859   &                   \\
1.5    & 0.48645575356     &                  &  3.28033126465   &                   \\
2.0    & $-$0.07571601608  & $-$0.07570\footnotemark[1],$-$0.07584\footnotemark[2]             
       &  1.62506812609    &                   \\ 
6.0    & $-$0.45051125895  & $-$0.45035\footnotemark[1],$-$0.45053\footnotemark[2]$^,$\footnotemark[5],$-$0.45109\footnotemark[3], 
       & $-$0.0081227650   & $-$0.00812\footnotemark[1],$-$0.00815\footnotemark[2],$-$0.00294\footnotemark[3],   \\ 
       &                   & $-$0.44945\footnotemark[4]     
       &                   & $-$0.00808\footnotemark[4],$-$0.00865\footnotemark[5],$-$0.00782\footnotemark[6]    \\ 
8.0    & $-$0.45122399716  & $-$0.45118\footnotemark[1],$-$0.45122\footnotemark[2]$^,$\footnotemark[5],$-$0.45193\footnotemark[3],  
       & $-$0.05762842270  & $-$0.05762\footnotemark[1]$^,$\footnotemark[4],$-$0.05764\footnotemark[2],$-$0.05293\footnotemark[3],    \\ 
       &                   & $-$0.45076\footnotemark[4]    
       &                   & $-$0.05783\footnotemark[5],$-$0.05510\footnotemark[6]    \\ 
10.0   & $-$0.45124920877  & $-$0.45124\footnotemark[1],$-$0.45125\footnotemark[2]$^,$\footnotemark[5],$-$0.45179\footnotemark[3],  
       & $-$0.0724869919   & $-$0.07247\footnotemark[1],$-$0.07250\footnotemark[2],$-$0.07008\footnotemark[3],    \\ 
       &                   & $-$0.45098\footnotemark[4]   
       &                   & $-$0.07243\footnotemark[4],$-$0.07257\footnotemark[5],$-$0.07196\footnotemark[6]    \\ 
15.0   & $-$0.45124999990  & $-$0.45125\footnotemark[1]$^,$\footnotemark[2]      
       & $-$0.07888401652  & $-$0.07885\footnotemark[1],$-$0.07888\footnotemark[2]                   \\  
25.0   & $-$0.45125000000  & $-$0.45125\footnotemark[1]$^,$\footnotemark[2]$^,$\footnotemark[4]$^,$\footnotemark[5],$-$0.45131\footnotemark[3]
       & $-$0.07917921743  & $-$0.07916\footnotemark[1],$-$0.07918\footnotemark[2]$^,$\footnotemark[5],$-$0.07920\footnotemark[3],    \\
       &                   &
       &                   & $-$0.07915\footnotemark[4],$-$0.07921\footnotemark[6]    \\ 
40.0   & $-$0.45124999999  &      
       & $-$0.07917943910  & $-$0.07918\footnotemark[1]$^,$\footnotemark[2]                   \\  
50.0   & $-$0.45124999999  & $-$0.45126\footnotemark[3],$-$0.45125\footnotemark[4]$^,$\footnotemark[5] 
       & $-$0.07917943910  & $-$0.07918\footnotemark[1]$^,$\footnotemark[2]$^,$\footnotemark[4]$^,$\footnotemark[5],$-$0.07920\footnotemark[3]$^,$\footnotemark[6]   \\
\end{tabular}
\end{ruledtabular}
\begin{tabbing}
$^{\mathrm{a}}$Ref.~\cite{varshni04}. \hspace{10pt}  \=
$^{\mathrm{b}}$Exact result, quoted in \cite{varshni04}. \hspace{10pt}  \=
$^{\mathrm{c}}$Ref.~\cite{sinha00}. \hspace{10pt}  \=
$^{\mathrm{d}}$Ref.~\cite{filho03}. \hspace{10pt}  \=
$^{\mathrm{e}}$Exact result, quoted in \cite{filho03}. \hspace{10pt}  \=
$^{\mathrm{f}}$Ref.~\cite{sinha03}.  \=
\end{tabbing}
\end{table}
\endgroup

Tables VII and VIII give sample eigenvalues for two low-lying nodeless states corresponding to $\ell=0,1$, \emph{viz.,} $1s, 2p$ of confined 
Hulth\'{e}n potential for $\delta=0.1$ and 0.2 respectively. In both cases, position of the spherical wall was selected at 12 
different locations, so as to cover \emph{small, intermediate and large} range of confinement. The lowest cavity radius so far considered in 
literature is: $r_c < 2$ (in case of $1s$, for both $\delta$) and 6 (in case of $2p$, for both $\delta$). Current energies show decent agreement 
with super-symmetric result of \cite{varshni04} for all values of box size for both states, wherever those are available. The same author 
also estimated these states by a numerical method that employed Numerov's method with a logarithmic mesh for solution of Schr\"odinger equation. 
The latter shows slightly better agreement than super-symmetric result, especially in the neighborhood of \emph{critical cage radius}, 
$r_c^c$. This is defined as the radius of the enclosure at which energy becomes zero, analogous to CHA. We have not attempted a detailed study. 
Rather a more restrictive approach is adopted by estimating a few limited ones. Thus, numerically obtained values of these for $1s, 2p$ states 
are: 1.8639458, 1.8939725, 1.9584319 and 5.4189704, 5.8257603, 7.0428492 respectively, for $\delta=0.05, 0.1$ and 0.2. These are in good accord with 
reported values of 1.894 ($\delta=0.1$), 1.958 ($\delta=0.2$) and 5.826 ($\delta=0.1$), 7.043 ($\delta=0.2$) for the same states
corresponding to screening parameters given in parentheses \cite{varshni04}. In another estimate \cite{tian06}, $r_c$ values are found to be
1.894, 1.958 (for $1s$ state having $\delta=0.1$, 0.2 respectively), whereas for $2p$ state, the corresponding values are 5.826 and 7.043. Excepting 
$\delta=0.2$ of $1s$, other three cases were studied by $1/N$ expansion method \cite{sinha00}. One notices that, generally, as 
$r_c$ goes to higher values, these results tend to improve. For same three cases, energies were reported by super-symmetric variational 
method \cite{filho03} and WKB method \cite{sinha03}, leading to quite similar accuracy and conclusions as those in \cite{sinha00}. 
Numerical estimates, as quoted in \cite{filho03}, are also produced wherever possible. While these reference energies show good agreement with 
each other, present values are considerably more accurate than all of these.

\begingroup
\squeezetable
\begin{table}
\caption {\label{tab:table8} Comparison of some low-lying states of confined Hulth\'{e}n potential with literature data, for $\delta=0.2$. 
PR implies Present Result.} 
\begin{ruledtabular}
\begin{tabular}{l|ll|ll}
$r_c$  & E$_{1s}$(PR)      & E$_{1s}$(Ref.) &      E$_{2p}$(PR)   &  E$_{2p}$(Ref.)   \\
\hline
0.1    & 469.092872952     &                &   991.107392541   &                   \\ 
0.5    & 14.8471617711     &                &  36.7578980396    &                   \\
1.0    & 2.4724301161      &                &  8.32120019501    &                   \\
1.5    & 0.53476949305     &                &  3.32817211249    &                   \\
2.0    & $-$0.02786272697  & $-$0.02784\footnotemark[1],$-$0.02800\footnotemark[2]                
       &  1.67221901970    &                                                        \\ 
6.0    & $-$0.40421171842  & $-$0.40404\footnotemark[1],$-$0.40423\footnotemark[2]        
       &  0.03422237169    &                                                        \\ 
8.0    & $-$0.40497046759  & $-$0.40493\footnotemark[1],$-$0.40497\footnotemark[2]                
       & $-$0.01709196413  & $-$0.01709\footnotemark[1],$-$0.01710\footnotemark[2],$-$0.01242\footnotemark[3],    \\ 
       &                   &                
       &                   & $-$0.01708\footnotemark[4],$-$0.01607\footnotemark[5],$-$0.01731\footnotemark[6]    \\ 
10.0   & $-$0.40499902640  & $-$0.40499\footnotemark[1],$-$0.40500\footnotemark[2]      
       & $-$0.0332989638   & $-$0.03329\footnotemark[1],$-$0.03330\footnotemark[2],$-$0.03118\footnotemark[3],    \\ 
       &                   &                         
       &                   & $-$0.03323\footnotemark[4],$-$0.03389\footnotemark[5],$-$0.03339\footnotemark[6]    \\ 
15.0   & $-$0.40499999985  & $-$0.40499\footnotemark[1],$-$0.40500\footnotemark[2]
       & $-$0.04128265021  & $-$0.04125\footnotemark[1],$-$0.04128\footnotemark[2]                               \\  
25.0   & $-$0.40499999999  & $-$0.40500\footnotemark[1]$^,$\footnotemark[2]     
       & $-$0.04188395482  & $-$0.04188\footnotemark[1]$^,$\footnotemark[2]$^,$\footnotemark[6],$-$0.04199\footnotemark[3],    \\     
       &                   &               
       &                   & $-$0.04178\footnotemark[4],$-$0.04192\footnotemark[5]    \\ 
40.0   & $-$0.40499999999  &          
       & $-$0.04188604888  & $-$0.04188\footnotemark[1],$-$0.04189\footnotemark[2]                          \\  
50.0   & $-$0.40499999999  &          
       & $-$0.04188604921  & $-$0.04189\footnotemark[1]$^,$\footnotemark[2]$^,$\footnotemark[4]$^,$\footnotemark[6],$-$0.04196\footnotemark[3],
         $-$0.04191\footnotemark[5]  \\
\end{tabular}
\end{ruledtabular}
\begin{tabbing}
$^{\mathrm{a}}$Ref.~\cite{varshni04}. \hspace{10pt}  \=
$^{\mathrm{b}}$Exact result, quoted in \cite{varshni04}. \hspace{10pt}  \=
$^{\mathrm{c}}$Ref.~\cite{sinha00}. \hspace{10pt}  \=
$^{\mathrm{d}}$Ref.~\cite{filho03}. \hspace{10pt}  \=
$^{\mathrm{e}}$Ref.~\cite{sinha03}. \hspace{10pt}  \=
$^{\mathrm{f}}$Exact result, quoted in \cite{filho03}. \=
\end{tabbing}
\end{table}
\endgroup

Next, Table IX presents calculated energies of Hulth\'en potential under spherical confinement for some representative moderately high-lying states. 
As an illustration, the screening parameter is fixed at $\delta=0.05$ and seven states are chosen corresponding to $n=3,4$, at ten selected values 
of $r_c$, \emph{viz.}, 0.1, 0.5, 1, 2, 5, 10, 20, 30, 50 and 100 a.u., respectively. In each case, energy, much like the case of CHA, steadily 
decreases from a high positive value to attain a negative value for a sufficiently high $r_c$ and remains stationary thereafter. Full confinement 
region is scanned. To the best of our knowledge, no such attempt is known for such states and they may offer a useful set of reference for 
future works in this direction.  

\begingroup
\squeezetable
\begin{table}
\caption {\label{tab:table9} Eigenvalues (a.u.) of $n=3,4$ states of confined Hulth\'{e}n potential for $\delta=0.05$.} 
\begin{ruledtabular}
\begin{tabular}{clllll}
State  & $r_c=0.1$      &   $r_c=0.5$        &      $r_c=1$     &       $r_c=2$     &   $r_c=5$         \\
\hline
$3s$   & 4406.1466414   & 170.61011193       & 40.888019695     & 9.3389386065      & 1.0776696979      \\
$3p$   & 2960.4872910   & 114.66849642       & 27.498883279     & 6.2937792870      & 0.7321605002      \\
$3d$   & 1644.5549089   & 63.185117264       & 14.992330190     & 3.3522434990      & 0.3534709790      \\
$4s$   & 7857.6541745   & 308.22219525       & 75.155388308     & 17.840882601      & 2.4067849721      \\
$4p$   & 5918.2078780   & 232.45290563       & 56.782925033     & 13.535366380      & 1.8548769549      \\
$4d$   & 4115.6076200   & 161.38194690       & 39.340200641     & 9.3389126846      & 1.2640613859      \\
$4f$   & 2426.4205347   & 94.651526411       & 22.920683088     & 5.3668108085      & 0.6937529695      \\
\hline
       & $r_c=10$       &   $r_c=20$         &      $r_c=30$    &   $r_c=50$       &  $r_c=100$        \\
\hline
$3s$   & 0.1152515762   & $-$0.0272469617    & $-$0.0331906519  & $-$0.033368031   &  $-$0.033368055   \\
$3p$   & 0.0730498617   & $-$0.0288228707    & $-$0.0330474388  & $-$0.033164486   &  $-$0.033164501   \\
$3d$   & 0.0166906839   & $-$0.0309487936    & $-$0.0327122864  & $-$0.032753179   &  $-$0.032753184   \\
$4s$   & 0.4290276270   & 0.0393206209       & $-$0.003319140   & $-$0.011126878   &  $-$0.011249999   \\
$4p$   & 0.3399459526   & 0.0307423709       & $-$0.0045189174  & $-$0.010961108   &  $-$0.011058170   \\
$4d$   & 0.2262751307   & 0.0171189105       & $-$0.0064514215  & $-$0.010611090   &  $-$0.010667404   \\
$4f$   & 0.1118826554   & 0.0024927744       & $-$0.0083336240  & $-$0.010043093   &  $-$0.010061964   \\
\end{tabular}
\end{ruledtabular}
\end{table}
\endgroup

We graphically show the effect of isotropic compression on energies of Hulth\'en potential in Fig.~2 now. In left side (a), these are 
given for all six states corresponding to $n=1-3$ having, $\delta=0.2$, while right side (b) considers all nine states 
belonging to $n=4,5$ with screening parameter 0.05 respectively. In both occasions, positive and negative energies are included. Shapes of 
all these curves are quite similar to each other. Note the axes of energy and box radius are different in two cases. Confinement within very 
small box size is again avoided for easy appreciation of figures. They all exhibit a sharp increase as $r_c$ becomes smaller. Plots 
remain well separated at relatively smaller $r_c$. From an initial high positive value, they fall off rapidly monotonically as $r_c$ increases, 
finally approaching the energy of corresponding free system smoothly and remaining constant thereafter. As $r_c$ decreases, energies change 
sign from negative to positive values becoming zero for critical cavity radius. Like the case of Coulomb potential, for a given $\delta$ and
$r_c$, state with \emph{lowest} $n$ remains lowest in energy within a particular $\ell$, whereas, for a given $n$, state with \emph{highest} $\ell$   
corresponds to lowest energy. Furthermore, for a specific $\delta$, sequence of energy follows the \emph{same} pattern as Coulomb potential of previous 
section in the limit of $r_c \rightarrow 0$. However, the same in unconfined case is as follows:
\[ 1s,2s,2p,3s,3p,3d,4s,4p,4d,4f,5s,5p,5d,5f,5g,6s,6p,6d,6f,6g,  \cdots \]

\begin{figure}
\centering
\begin{minipage}[b]{0.40\textwidth}\centering
\includegraphics[scale=0.38]{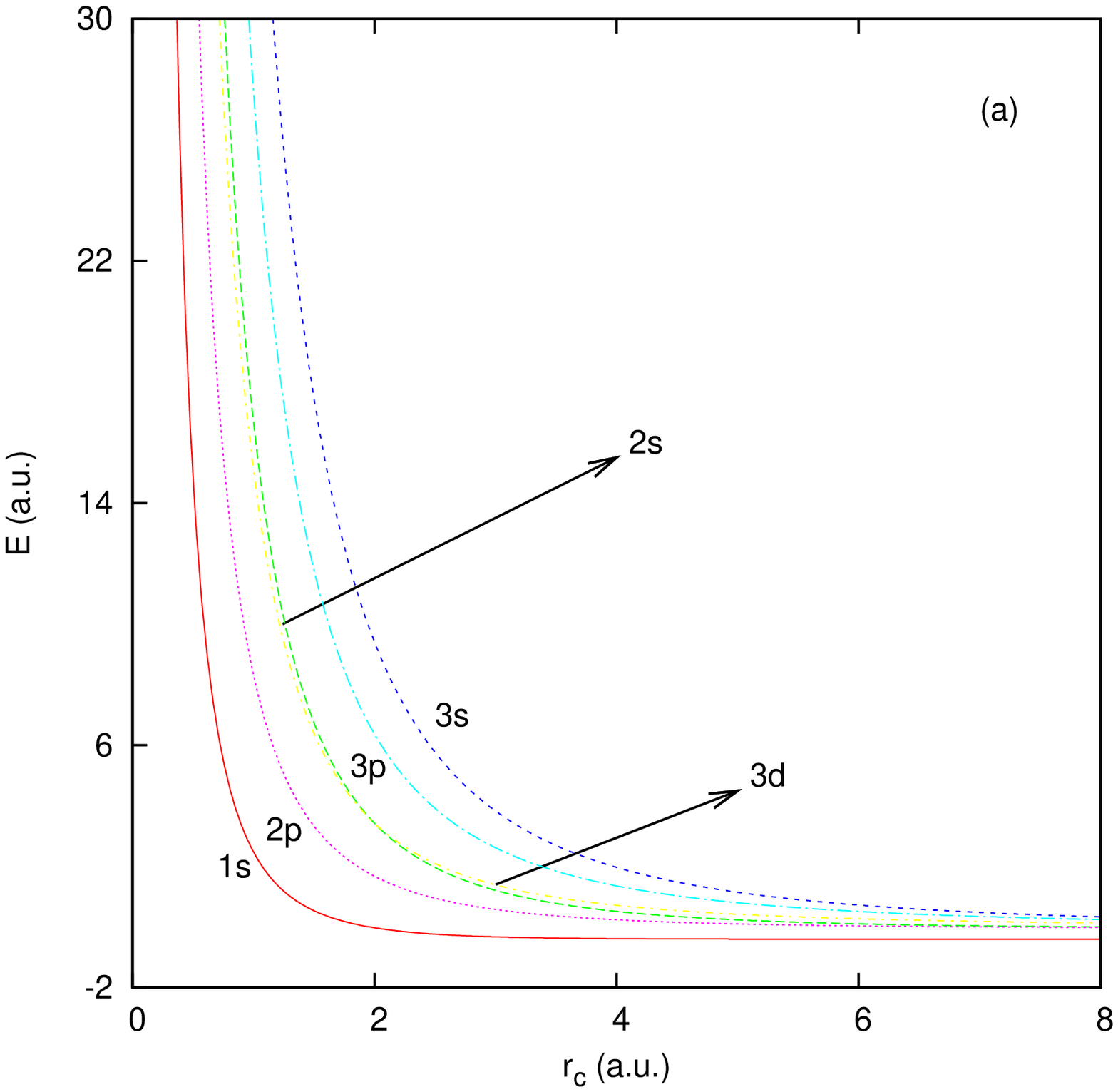}
\end{minipage}
\hspace{0.15in}
\begin{minipage}[b]{0.35\textwidth}\centering
\includegraphics[scale=0.38]{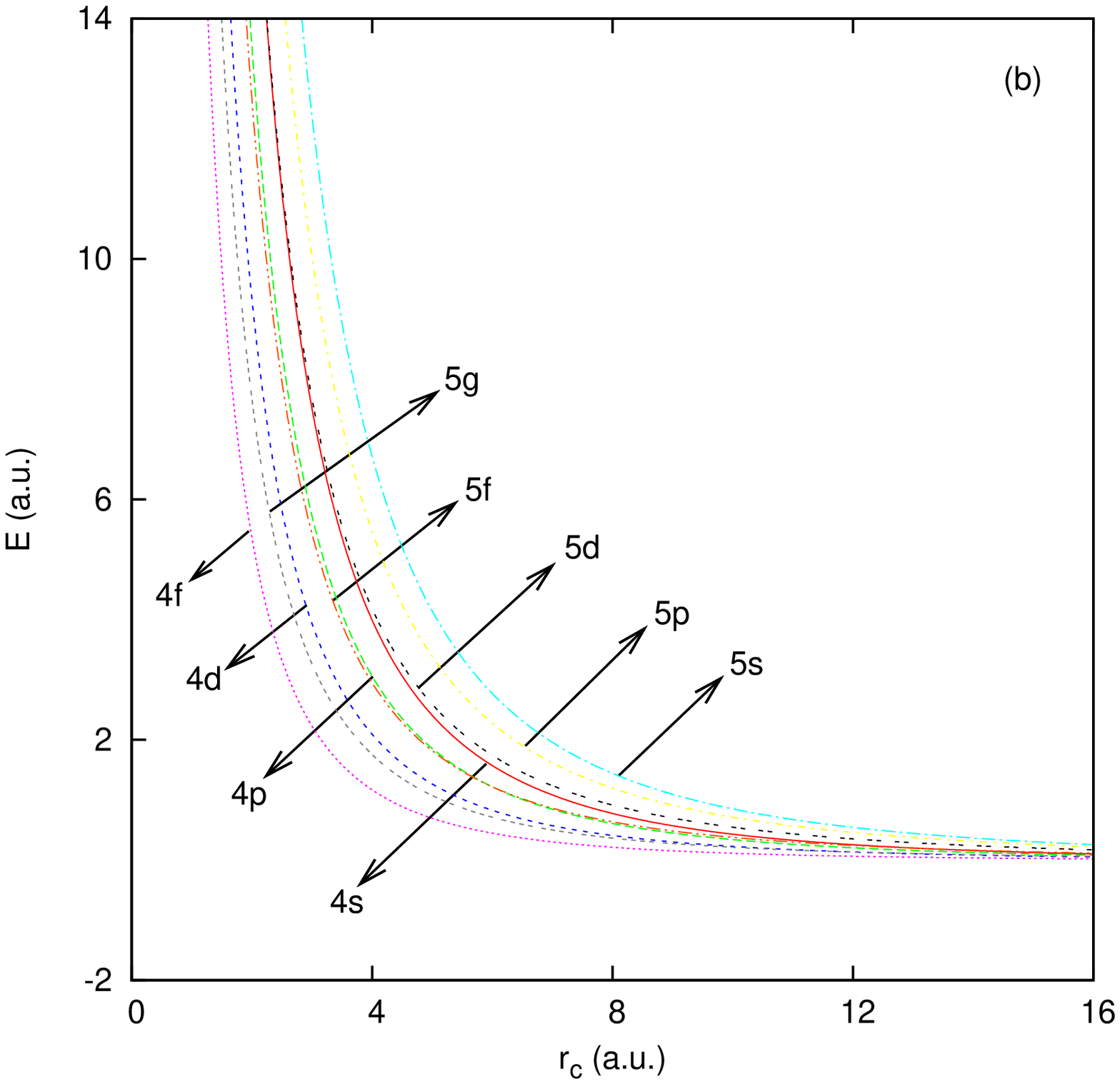}
\end{minipage}
\caption[optional]{Energy variations in Hulth\'{e}n potential with confinement radius: (a) $n=1,2,3$ (b) $n=4,5$. Corresponding $\delta$
values are 0.2 and 0.05 for (a), (b). See text for details.}
\end{figure}

Finally, a few words about the dipole polarizability of confined Hulth\'en potential. The exact calculation of polarizability is 
quite involved; we use simplified expressions, originally derived for one-electron atoms in free-space, by Kirkwood \cite{kirkwood32} 
and Buckingham \cite{buckingham37}, namely, 
\begin{equation}
\alpha_D^K= \frac{4}{9} \langle r^2 \rangle^2; \ \ \ \ \ 
\alpha_D^B= \frac{2}{3} \left[ 
\frac{6 \langle r^2 \rangle^3 +3 \langle r^3 \rangle^2 - 8 \langle r \rangle \langle r^2 \rangle \langle r^3 \rangle} 
{9 \langle r^2 \rangle - 8 \langle r \rangle^2 }         \right]
\end{equation}
Assuming that these expressions hold good for confined systems, as has been done in many previous occasions, we summarize our results in Table X. 
Since a number of high-quality estimates are available for CHA problem (see, for example, \cite{aquino95,dutt01,montgomery02,aquino07}), we do not 
attempt those here and restrict ourselves to the Hulth\'en potential case. Thus, $\alpha_D^K$ and $\alpha_D^B$ are offered for $1s$ and $2p$ 
states corresponding to $\delta=0.1$, 0.2 respectively. In both states, $r_c$ values are chosen so as to cover a broad range. Some results
are reported \cite{varshni04} for ground state, which are duly quoted for comparison. No results could be found for excited state. For a 
given $\delta$, both $\alpha_D^K$ and $\alpha_D^B$ gradually increase with $r_c$, finally reaching an asymptotic value. For a given $\delta$, 
one finds that, $\alpha_D^K \leq \alpha_D^B$. Although the inequality holds for free H atom, there is no proof that the same is valid for CHA or
a confined Hulth\'en potential. As $r_c$ increases, difference between the two $\alpha$ tends to increase significantly. It is generally found that 
polarizability values for $2p$ states are much higher compared to the ground state for a given $\delta$; moreover, the asymptotic value is 
reached for a considerably large $r_c$ for $2p$ state. Furthermore, one notices an increase in both $\alpha_D^K$ and $\alpha_D^B$ values with an 
increase in $\delta$. Present results are significantly improved over the previous ones. 

\begingroup
\squeezetable
\begin{table}
\caption {\label{tab:table10} Dipole polarizability (in a.u.) of confined Hulth\'en potential with respect to cage radius.
Numbers in the parentheses denote reference values, taken from \cite{varshni04}.} 
\begin{ruledtabular}
\begin{tabular}{cc|ll|ll}
State  &  $r_c$   & \multicolumn{2}{c}{$\delta=0.1$}    &  \multicolumn{2}{c}{$\delta=0.2$}   \\
\cline{3-6} 
       &          &   $\alpha_D^K$        &  $\alpha_D^B$        &  $\alpha_D^K$         &   $\alpha_D^B$         \\
\hline
$1s$   & 0.5      & 0.002002769           &  0.002030037         &  0.002002790          &  0.002030059           \\ 
       & 1.0      & 0.028478053           &  0.028675462         &  0.028480610          &  0.028678122           \\ 
       & 1.5      & 0.125863099           &  0.126135536         &  0.125903929          &  0.126177170           \\ 
       & 2.0      & 0.340234836(0.340)    &  0.340249553(0.340)  &  0.340512990(0.340)   &  0.340528358(0.340)    \\
       & 3.0      & 1.17445441(1.175)     &  1.18214819(1.183)   &  1.17798124(1.180)    &  1.18562199(1.188)     \\
       & 4.0      & 2.29626217(2.29)      &  2.36328371(2.35)    &  2.31263673(2.31)     &  2.37966765(2.37)      \\    
       & 5.0      & 3.21734552(3.18)      &  3.41693628(3.36)    &  3.25748090(3.24)     &  3.45908008(3.41)      \\ 
       & 6.0      & 3.72673722(3.67)      &  4.07085354(3.97)    &  3.79120326(3.77)     &  4.14280842(4.08)      \\ 
       & 8.0      & 4.00387282(3.93)      &  4.48772555(4.36)    &  4.09197277(4.06)     &  4.59462306(4.51)      \\ 
       & 10.0     & 4.02860091(4.00)      &  4.53399268(4.48)    &  4.12069600(4.09)     &  4.64835736(4.59)      \\    
\hline
$2p$   & 0.5      & 0.003785194           &  0.004204328         &  0.003785213          &  0.004204351           \\
%      & 1.0      & 0.058714724           &  0.064875982         &  0.058717104          &  0.064878951           \\ 
%      & 2.0      & 0.878277396           &  0.959982765         &  0.878582799          &  0.960356595           \\
       & 3.0      & 4.12373508            &  4.45692552          &  4.12892415           &  4.46314848            \\
%      & 4.0      & 11.9770471            &  12.7964019          &  12.0153203           &  12.8413122            \\
       & 5.0      & 26.5905910            &  28.0822696          &  26.7681022           &  28.2858721            \\
%      & 6.0      & 49.5463611            &  51.7344874          &  50.1560201           &  52.4176237            \\
%      & 7.0      & 81.3888796            &  84.0725162          &  83.0781424           &  85.9218331            \\
       & 8.0      & 121.326555            &  124.109261          &  125.295494           &  128.358067            \\
%      & 9.0      & 167.204140            &  169.620908          &  175.360315           &  178.174603            \\
       & 10.0     & 215.812725            &  217.508104          &  230.776712           &  232.925530            \\
       & 15.0     & 395.035988            &  395.743476          &  482.306562           &  483.118500            \\
       & 20.0     & 434.150727            &  437.616438          &  578.351678           &  586.056324            \\
       & 25.0     & 437.322406            &  441.343442          &  595.270889           &  606.401640            \\
       & 30.0     & 437.477603            &  441.540161          &  597.272379           &  609.068736            \\
\end{tabular}
\end{ruledtabular}
\end{table}
\endgroup

\section{conclusion}
Accurate eigenfunctions, energies, radial expectation values are reported for two simple Coulombic systems, namely, H atom and Hulth\'en 
potential confined at the center of an impenetrable spherical cavity of radius $r_c$. The GPS procedure employed here, is able to offer 
high-quality results (energies correct up to eleven decimal place) uniformly for the \emph{entire} (small, intermediate and large) ranges 
of confinement. Results for low and higher 
states are obtained with equal ease and efficiency without necessitating any extensive algebraic manipulation, unlike some other methods. 
Effects of box radius on energy levels of enclosed system are examined systematically. Present results show, comparable agreement
with best theoretical estimates. Critical cavity radii for all states up to and including $n=10$ for CHA, have been examined. 
These have been studied by only few workers until now and could be helpful in future. For many states, previous results are significantly 
improved. The degeneracy breaking as well as energy ordering under the influence of confinement are also discussed. 
A similar kind of analysis, as for the CHA, has been made for the latter. In all cases, present results are noticeably superior
to all other existing values. Changes of critical screening parameter with respect to confinement radius is briefly discussed.
For the latter potential, accurate dipole polarizabilities are provided as well. Considering the simplicity and accuracy offered by this 
method, it may be also successful and useful for other confinement situations in quantum mechanics.  

\section{acknowledgment} Critical constructive comments and suggestions from two kind anonymous referees and the Editor have significantly 
improved the manuscript. I sincerely thank Prof.~N.~A.~Aquino for kindly supplying a copy of the reference \cite{aquino14}. I am indebted to
Prof.~K.~D.~Sen for introducing me in this fascinating area. It is a pleasure to thank Prof.~Raja Shunmugam for extending valuable support. The 
help of our Librarian, Mr.~Siladitya Jana, is acknowledged, who procured a few references. Mr.~Shahid Ali Farooqui and Mr.~Sanjib Das is thanked 
for their assistance in the figure.

\end{document}